\documentclass[english]{llncs}
\usepackage{amsmath,amsfonts,amssymb,epsfig}
\newcommand{\argmin}[1]{\underset{#1}{\mbox{arg}\min}}
\usepackage{subfigure}
\usepackage{makeidx}  
\usepackage{url}      
\usepackage[latin1]{inputenc}
\usepackage{babel}

\usepackage{graphicx}
\usepackage{algorithm}
\usepackage{algpseudocode}
\usepackage{array}
\begin{document}
\frontmatter          
\pagestyle{empty}  
\mainmatter              
\title{Prostate Biopsy Assistance System with Gland Deformation Estimation for Enhanced Precision}
\titlerunning{Prostate Biopsy Assistance}  
\author{Michael Baumann\inst{1,3} \and Pierre Mozer\inst{2} \and Vincent Daanen\inst{3} \and Jocelyne Troccaz\inst{1}}
\authorrunning{Baumann et al.}   
\tocauthor{M. Baumann (University Joseph Fourier),
P. Mozer (Pitié-Salpétrière Hospital),
V. Daanen (Koelis SAS),
J. Troccaz (University Joseph Fourier)
}
\institute{Université J.Fourier, TIMC laboratory, Grenoble, France; CNRS, UMR 5525; 
  \and Koelis SAS, 5. av. du Grand Sablon, 38700 La Tronche, France;
  \and La Pitié-Salpêtrière hospital, urology dpt, 75651 Paris Cedex 13, France.
  \\ \email{michael.baumann@imag.fr}
  \thanks{Thanks to the Agence Nationale de la Recherche (TecSan project, France), the French Ministry of Health (PHRC program, France) and to Koelis S.A.S. (France) for funding.}
}

\maketitle              

\begin{abstract}
Computer-assisted prostate biopsies became a very active research area during the last years. Prostate tracking makes it possible to overcome several drawbacks of the current standard transrectal ultrasound (TRUS) biopsy procedure, namely the insufficient targeting accuracy which may lead to a biopsy distribution of poor quality, the very approximate knowledge about the actual location of the sampled tissues which makes it difficult to implement focal therapy strategies based on biopsy results, and finally the difficulty to precisely reach non-ultrasound (US) targets stemming from different modalities, statistical atlases or previous biopsy series. The prostate tracking systems presented so far are limited to rigid transformation tracking. However, the gland can get considerably deformed during the intervention because of US probe pressure and patient movements. We propose to use 3D US combined with image-based elastic registration to estimate these deformations. A fast elastic registration algorithm that copes with the frequently occurring US shadows is presented. A patient cohort study was performed, which yielded a statistically significant in-vivo accuracy of 0.83$\pm$0.54mm.
\end{abstract}

\section{Introduction}
\label{sec:intro}
Prostate biopsies are the only definitive way to confirm a prostate cancer hypothesis. The current clinical standard is to perform prostate biopsies under 2D TRUS control. The US probe is equipped with a needle guide for transrectal access of the prostate. The guide aligns the needle trajectory with the US image plane, which makes it possible to visualize the trajectory on the image for needle placement control. Unfortunately, in particular mid- and early-stage carcinoma are mostly isoechogenic, i.e. not visible in US images, which makes it necessary to sample the gland according to a systematic pattern. It is common to acquire 10 to 12 systematically distributed biopsies, the standard pattern taking roughly into account that most tumors (70$\%$) develop in the peripheral zone of the gland.

The current standard biopsy procedure has several shortcomings: first, it is difficult for the clinician to reach systematic targets accurately because he has to move the probe continuously to place the needle; a constant visual reference is hence lacking. Second, performing a non-exhaustive systematic search for an invisible target implies that the target can be missed. Negative results leave the clinician in a dilemma when the cancer hypothesis cannot be discarded: his only option is to repeat the biopsy series. Furthermore, the location of the acquired samples with respect to the patient anatomy is only very approximately known after the intervention. Uncertainty about tumor location is the principal reason why prostate therapy is in general radical. 

In order to address these issues, Baumann et al. \cite{baumann07tracking} and Xu et al. \cite{xu07protrack} simultaneously proposed to acquire a US volume before the intervention and to use it as anatomical reference. The stream of US control images acquired during the intervention is then registered with the reference volume, which allows to project targets defined in the reference volume into the control images, and, conversely, the biopsy trajectory, known in control image space, into the reference volume. This technique makes it possible to improve biopsy distribution accuracy by showing the current trajectory in a fixed reference together with the trajectories of previously acquired biopsies, to aim targets defined in the reference volume during a planning phase, and to know the precise biopsy positions after the intervention. Non-US targets could originate from suspicious lesions in MR volumes that are then multi-modally registered with the US reference volume. It is also possible to derive targets from more sophisticated statistical atlases \cite{shen04atlas} or, in the case of repeated biopsies, they could consist of previously unsampled regions. After the intervention, the biopsy trajectories in the reference volume can be combined with the histological results and used for therapy planning. 

Xu et al. \cite{xu07protrack} acquire a freehand 3D US volume and use 2D control images during the intervention. The 2D control images are tracked in operating room space with a magnetic sensor on the probe. In a second step, image-based registration is performed to compensate for small organ and patient movements. A similar approach was proposed by Bax et al. \cite{bax08biopsyguidance}, who use an articulated arm for 2D US beam tracking. Bax  does not compensate for patient and gland movements. However, pain-related pelvis movements are frequent, since the patient is not under total anesthesia. In that case, both methods risk to loose track of the gland because the US beam is tracked in operating room space and not in organ space, and a new reference volume has to be acquired. Baumann et al. address this draw-back by using 3D US to obtain richer control images during the intervention \cite{baumann07tracking}. Instead of using a US beam tracking device to initialize local image-based registration, they propose a kinematic model of endorectal probe movements to compute anatomically plausible positions of the US beam with respect to the gland, which is unaffected by patient movements.

However, probe movements during needle placement continuously deform the gland. Deformations are strongest near the probe head and can reach 3 to 6 mm. They cannot be estimated with the presented systems. To address this issue, we extend the 3D US rigid registration approach presented in \cite{baumann07tracking} by adding a deformation estimation step to the registration pipeline. 3D US control images provide the information required to estimate the deformation with acceptable precision and accuracy. Inverse consistency and linear elasticity are used as deformation priors. A novel image distance measure capable of dealing with local intensity shifts, frequent in US images, is presented. The clinical accuracy of the presented algorithm is evaluated on a large number of patient data acquired during prostate biopsy sessions.

\section{Method}

\begin{figure}
\centering
\includegraphics[width=0.9\textwidth]{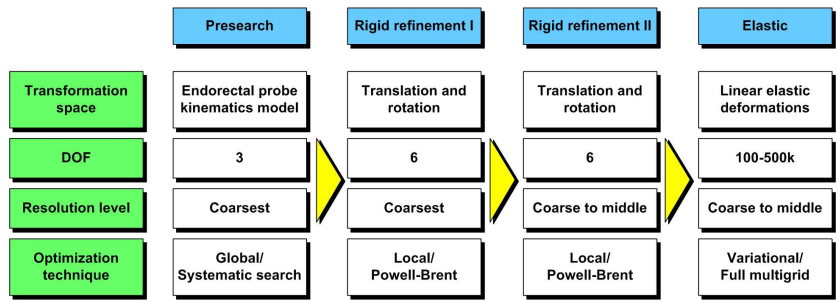}
\caption{Registration pipeline. The dimensionality of the transformation space and the image resolution are successively increased.}
\label{fig:pipeline}
\end{figure}

Non-linear registration of 3D US image streams is currently the most promising approach to perform organ tracking with deformation estimation. The principal challenges of image-based tracking systems are robustness and computational efficiency. A technique to achieve both goals are coarse-to-fine registration strategies that successively increase the degrees of freedom (DOF) of the transformation space and the image resolution. In this paper, we add an elastic registration step to the 3-step coarse-to-fine rigid registration pipeline that we proposed in \cite{baumann07tracking}. The resulting pipeline is illustrated in Fig.~\ref{fig:pipeline}.

\subsection{Framework for non-linear registration}

Image-based deformation estimation can be formulated as an optimization process of a local distance measure. Let $I_1, I_2:\mathbb{R}^3\rightarrow\mathbb{R}$ be images, $\varphi:\mathbb{R}^3\rightarrow\mathbb{R}^3$ the deformation function and the functional $\mathcal{D}[I_1,I_2;\varphi]$ a measure of the distance between $I_1$ and $I_2\circ\varphi$. In contrast to parametric approaches that use basis functions to build the deformation function, we will follow a variational approach and define $\varphi(x)=x+u(x)$, where $u:\mathbb{R}^3\rightarrow\mathbb{R}^3$ is assumed to be a diffeomorphism. The deformation could then be estimated by solving the optimization problem
\begin{equation}
\varphi^\ast=\argmin{\varphi}\left(\mathcal{E}[I_1,I_2;\varphi]\right),
\end{equation}
where the registration energy $\mathcal{E}$ simply corresponds to $\mathcal{D}$.
Straightforward minimization of a distance measure yields, however, in general poor results due to countless local minima, in particular in presence of noise, partial object occlusion and other imperfections in the image data. Unfortunately, US is a particularly noisy modality, which makes 3D US based deformation estimation vulnerable to local misregistrations. This problem can be addressed by integration of a priori models of the expected deformation. This can be done implicitly by adding further energy terms to the objective function. In this work, inverse consistency and elastic regularization energies are added. 

\subsection{Inverse Consistency Constraints}
In non-linear image registration, the forward estimation that minimizes $\mathcal{E}[I_1,I_2;\varphi]$ does in general not yield the inverse of the backward estimation that minimizes $\mathcal{E}[I_2,I_1;\psi]$, i.e. $\varphi \circ \psi \neq Id$ with $Id:\mathbb{R}^3\rightarrow\mathbb{R}^3,x\mapsto x$. Introduction of Zhang's inverse consistency constraint \cite{zhang06consistent} 
\begin{equation}
\mathcal{I}[\psi;\varphi]=\int_\Omega||\psi \circ \varphi - Id||_{\mathbb{R}^3}^2\ dx
\end{equation}
as additional energy penalizes solutions that lead to inconsistent inverse transformations, where $\Omega \subset \mathbb{R}^3$ is the registration domain in image space. Estimation of the forward and the backward deformations is coupled by an alternating iterative optimization
\begin{eqnarray}
\label{invconsistency}
\varphi^{k+1}=\argmin{\varphi}\left(\mathcal{E}[I_1,I_2;\varphi]+\mathcal{I}[\psi^{k};\varphi]\right),\\
\psi^{k+1}=\argmin{\psi}\left(\mathcal{E}[I_2,I_1;\psi]+\mathcal{I}[\varphi^{k};\psi]\right).
\end{eqnarray}
Concurrent estimation with mutual correction reduces the risk of local misregistrations. 

\subsection{Elastic Regularization}
The deformation of the prostate caused by probe pressure is fully elastic, which justifies the introduction of the linearized elastic potential \cite{Modersitzki04numerical} 
\begin{equation}
\mathcal{E}[\varphi]=\mathcal{E}[u+Id]=\int_\Omega \frac{\mu}{4}\sum^3_{j,k=1}\left(\partial_{x_j}u_k+\partial_{x_k}u_j\right)^2+\frac{\lambda}{2}(\mbox{div}\ u)^2\ dx
\end{equation}
as additional energy, where $\lambda$ and $\mu$ are the Lamé coefficients.

%
%
\subsection{Image Distance Measure}

The image distance measure is the driving energy of the optimization process. Experiments on patient data have shown that the sum of squared distances (SSD) is a poor distance measure for deformation estimation on noisy US images. Local intensity changes are frequent due to changing US beam angles with respect to the tissues and probe pressure variations. The more robust Pearson correlation coefficient (CC) requires the evaluation of a large neighborhood of every voxel pair to yield statistically significant results, which is incompatible with deep multi-resolution approaches that operate on very coarse levels. 

We hence prefer an intermediate correlation model that filters low-frequency intensity shifts, i.e. we assume that $I_1 \equiv I_2\circ \hat{\varphi} + b$, where $\hat{\varphi}$ is the physical solution of the registration problem, and where  $b:\mathbb{R}^3\rightarrow\mathbb{R}^3$ models a local intensity shift. The shift is estimated by
\begin{equation}
b^\sigma[\varphi](x)=(I_1-I_2\circ\varphi)\ast\mathcal{G}_\sigma(x)
\end{equation}
where $\mathcal{G}:\mathbb{R}^3\rightarrow\mathbb{R}$ is a Gaussian with standard deviation $\sigma$. The image distance energy is then
\begin{equation}
\mathcal{D}[I_1,I_2;\varphi]=\int_\Omega(I_1(x)-I_2(\varphi(x))-b^\sigma[\varphi](x))^2\ dx.
\end{equation}

The standard deviation $\sigma$ controls the frequency range of the high-pass filter. If $\sigma$ gets smaller, the cropped frequency range gets larger, and registration convergence rate decreases and may even stall if only high frequency noise like speckle is left. When used with a multi-resolution solver on a Gaussian pyramid (cf. next section), which implicitly performs a low-pass filtering of the intensity variations on coarse resolutions, this approach transforms to a band-pass filtering on varying frequency bands. In this configuration it is sufficient to chose relatively small standard deviations without risking registration inefficiencies.

\subsection{Solver}
Combination of the energy terms yields the alternating system
\begin{eqnarray}
\label{solver1}
\varphi^\ast &=&\argmin{\varphi}\left(\mathcal{D}[I_1,I_2,\varphi]+\mathcal{E}[\varphi]+\mathcal{I}[\psi;\varphi]\right),\\
\label{solver2}
\psi^\ast &=&\argmin{\psi}\left(\mathcal{D}[I_2,I_1,\psi]+\mathcal{E}[\psi]+\mathcal{I}[\varphi;\psi]\right).
\end{eqnarray}
An iterative two-step minimization scheme is used to solve both objective functions. The Euler-Lagrange equations of Eqn. \ref{solver1} and \ref{solver2} are rewritten as a fixed point iteration
\begin{eqnarray}
\label{partial1}
\frac{\varphi^{k+1}-\varphi^{k}}{\Delta t} = \mathcal{L}[\varphi^{k}] + f_\mathcal{D}[I_1,I_2;\varphi^{k}]+f_\mathcal{I}[\psi^{k};\varphi^{k}],\\
\label{partial2}
\frac{\psi^{k+1}-\psi^{k}}{\Delta t} = \mathcal{L}[\psi^{k}] +
f_\mathcal{D}[I_2,I_1;\varphi^{k}]+f_\mathcal{I}[\varphi^{k};\psi^{k}],
\end{eqnarray}
where $t \in \mathbb{R}$ controls the discretization granularity, and with the elliptic partial differential operator
\begin{equation}
\mathcal{L}[\varphi]=\mathcal{L}[u+Id]=\mu \Delta u + (\lambda + \mu)\nabla \mbox{div}\ u,
\end{equation}
which is obtained from the Gâteaux-derivative of $\mathcal{E}[\varphi]$\cite{Modersitzki04numerical}. The Gâteaux derivatives of the energy term $\mathcal{D}$ at $\varphi$ yields the force term
\begin{equation}
f_\mathcal{D}[I_1,I_2;\varphi]=\left(I_{1}-I_{2}-b^\sigma[\varphi]\ast\mathcal{G}_\sigma\right)\cdot(\nabla I_2)\circ\varphi.
\end{equation}
and for $\mathcal{I}$ we get
\begin{equation}
f_\mathcal{I}[\psi;\varphi]=(\nabla \psi)\circ\varphi\cdot(\psi\circ\varphi-Id).
\end{equation}

An iterative algorithm is used to estimate the displacement fields:
\begin{algorithmic}[1]
\While{not converged}
\State compute $f_\mathcal{D}[I_1,I_2;\varphi^{k}]$ and $f_\mathcal{I}[\psi^{k};\varphi^{k}]$
\State compute $f_\mathcal{D}[I_2,I_1;\psi^{k}]$ and $f_\mathcal{I}[\varphi^{k};\psi^{k}]$
\State solve Eqn. \ref{partial1} for $\varphi^{k+1}$
\State solve Eqn. \ref{partial2} for $\psi^{k+1}$
\EndWhile
\end{algorithmic}
The forces are hence considered as constants for the resolution of the PDEs \ref{partial1} and \ref{partial2}, and the forward and the backward estimation correct themselves mutually at each force update. The PDEs are solved using Red-Black Gauss-Seidel relaxation. Convergence is achieved if the difference of the $L_2$-norm of the total forces between two iterations is below a threshold for both the forward and the backward estimation (oscillatory states are detected). The algorithm is executed on various resolution levels of a Gaussian image pyramid \cite{baumann07tracking} using the full multigrid strategy \cite{briggs08multigrid}. Note that the algorithm derives from the multigrid scheme by iterating until convergence on every grid level. This is necessary since the relaxation is performed on fractional forces. Fixed edges and bending side walls are used as border conditions \cite{Modersitzki04numerical}. The elasticity parameters are chosen such that Poisson's coefficient is zero, hence maximizing compressibility to allow compensation of local model inadequacies. Young's modulus is interpreted as a free variable in function of Poisson's coefficient and the PDE discretization $\Delta t$ since it has no physical meaning in image registration. The forces are capped to a maximum length which makes it possible to control the maximum contributions per iteration to the displacement field via $\Delta t$. Limiting the contributions to less than 0.5 voxel side lengths ensures that the algorithm does not 'jump' over intensity barriers during optimization.

\section{Experiments}
\begin{figure}
\centering
\subfigure[]{\includegraphics[width=.245\textwidth]{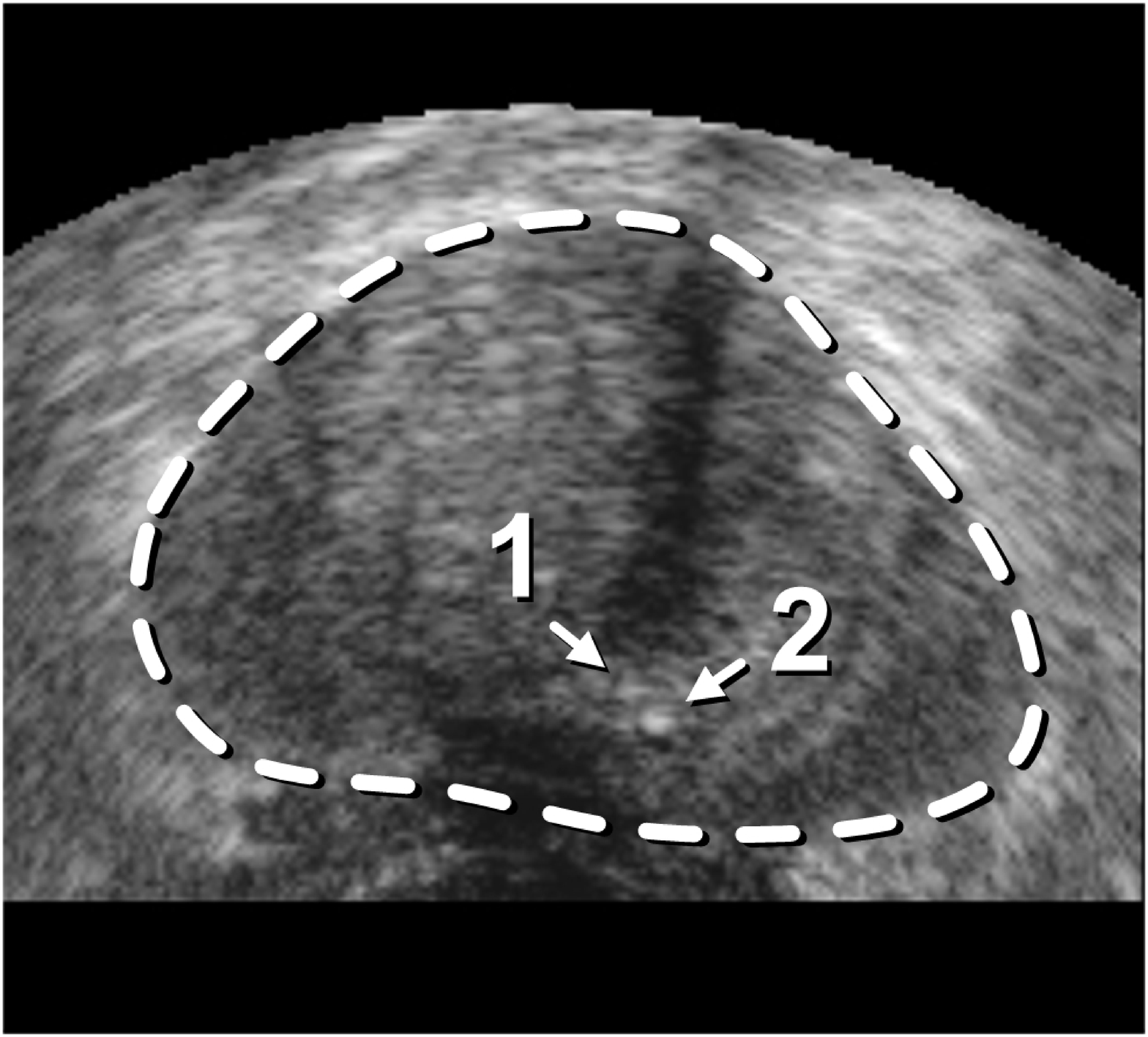}}\hfill
\subfigure[]{\includegraphics[width=.245\textwidth]{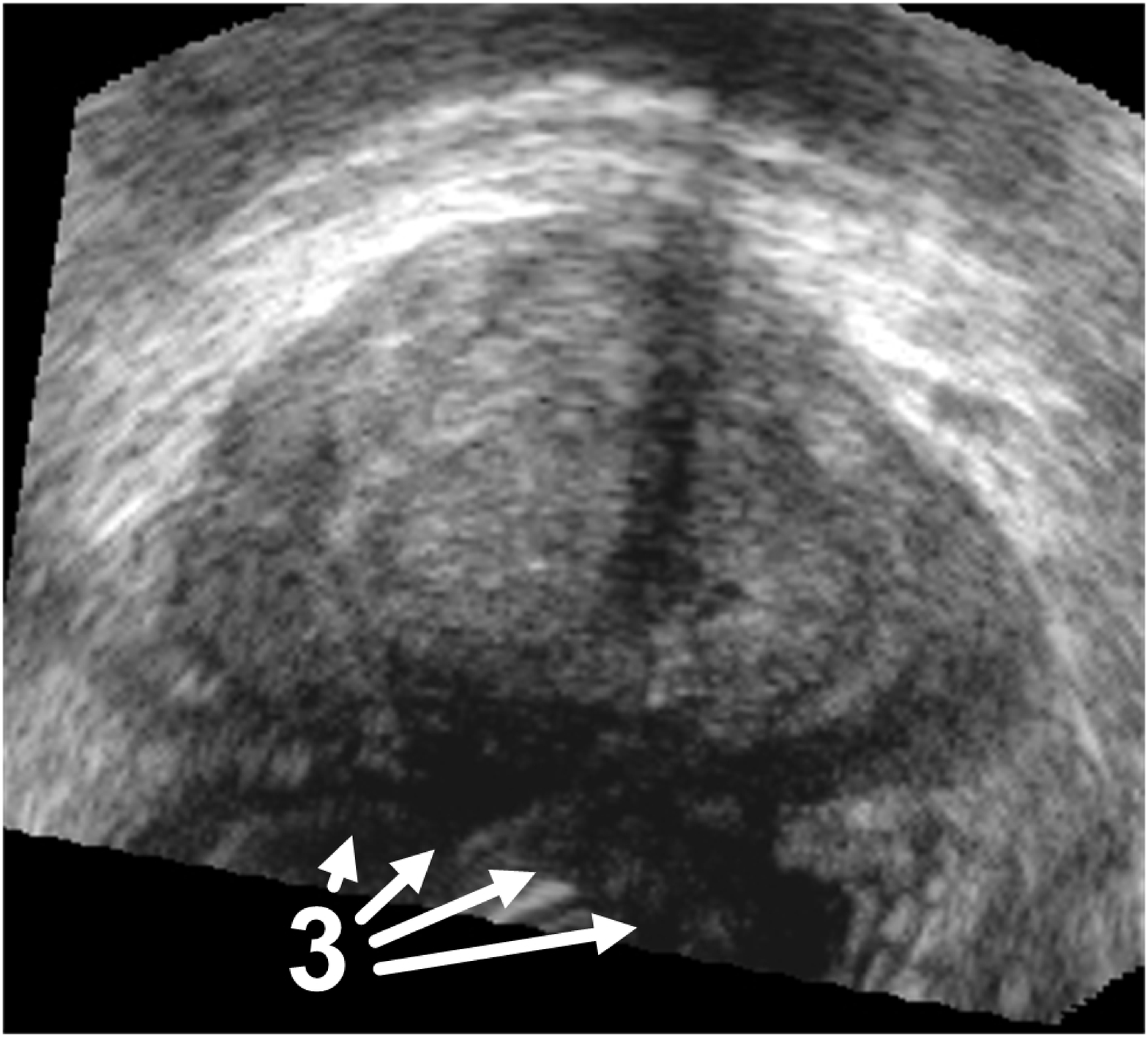}}\hfill
\subfigure[]{\includegraphics[width=.245\textwidth]{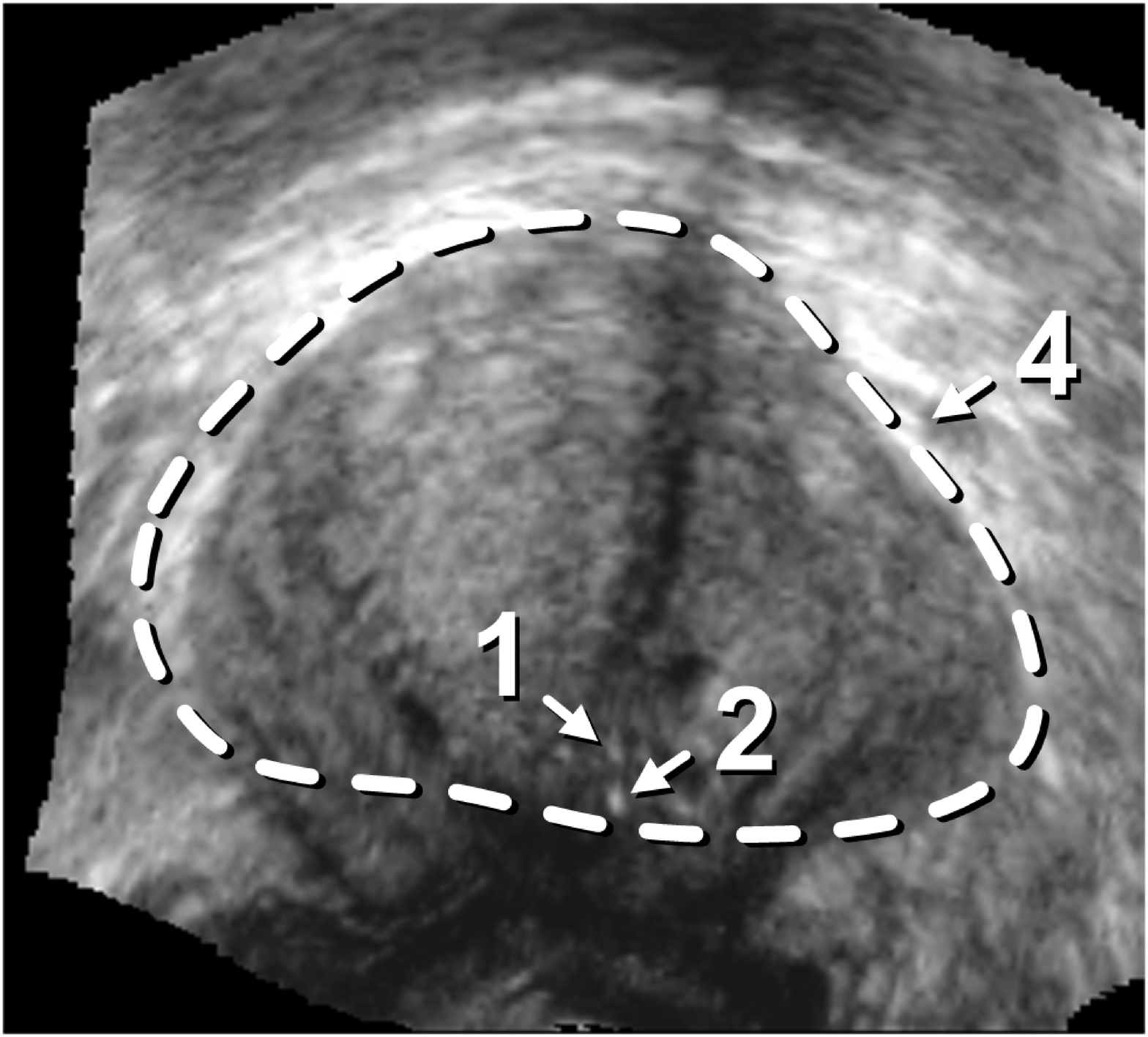}}\hfill
\subfigure[]{\includegraphics[width=.245\textwidth]{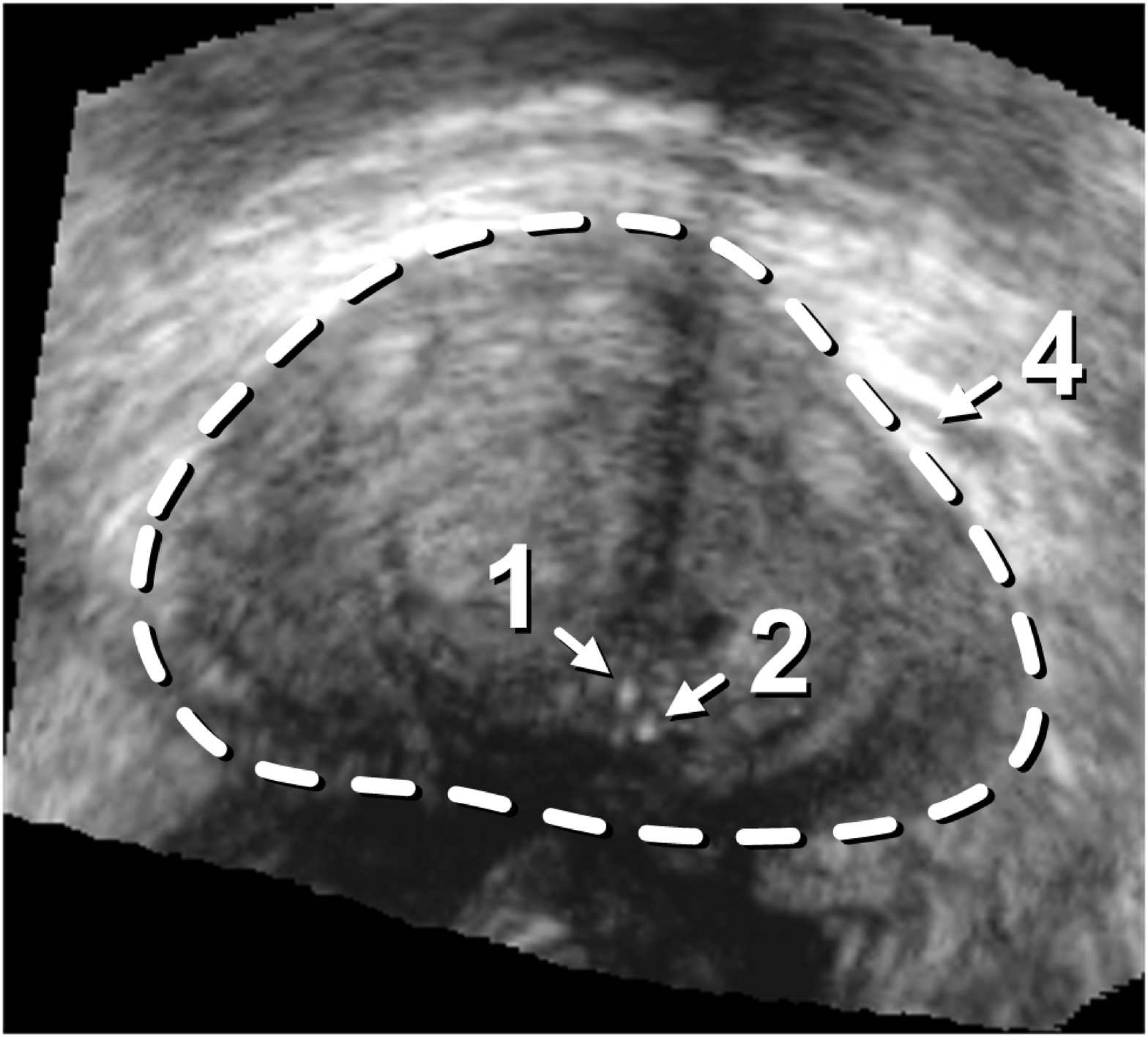}} \vspace{-4pt}
\caption{Fig. (a) shows a prostate volume with calcifications [1,2]. Fig. (b) shows a second volume after rigid registration; low probe pressure led to the low contrast zone [3]. Fig. (c) shows the 3D elastic registration with standard SSD; the whole prostate is dragged towards zone [3]. Fig. (d) shows the 3D intensity shift filtered, inverse consistent elastic registration; the strong intensity differences between both volumes are correctly handled, the calcifications make appearance at the correct position (best viewed in PDF with zoom).}
\label{fig:SSD}
\end{figure}

\begin{figure}
\centering
\subfigure[]{\includegraphics[width=0.3\textwidth]{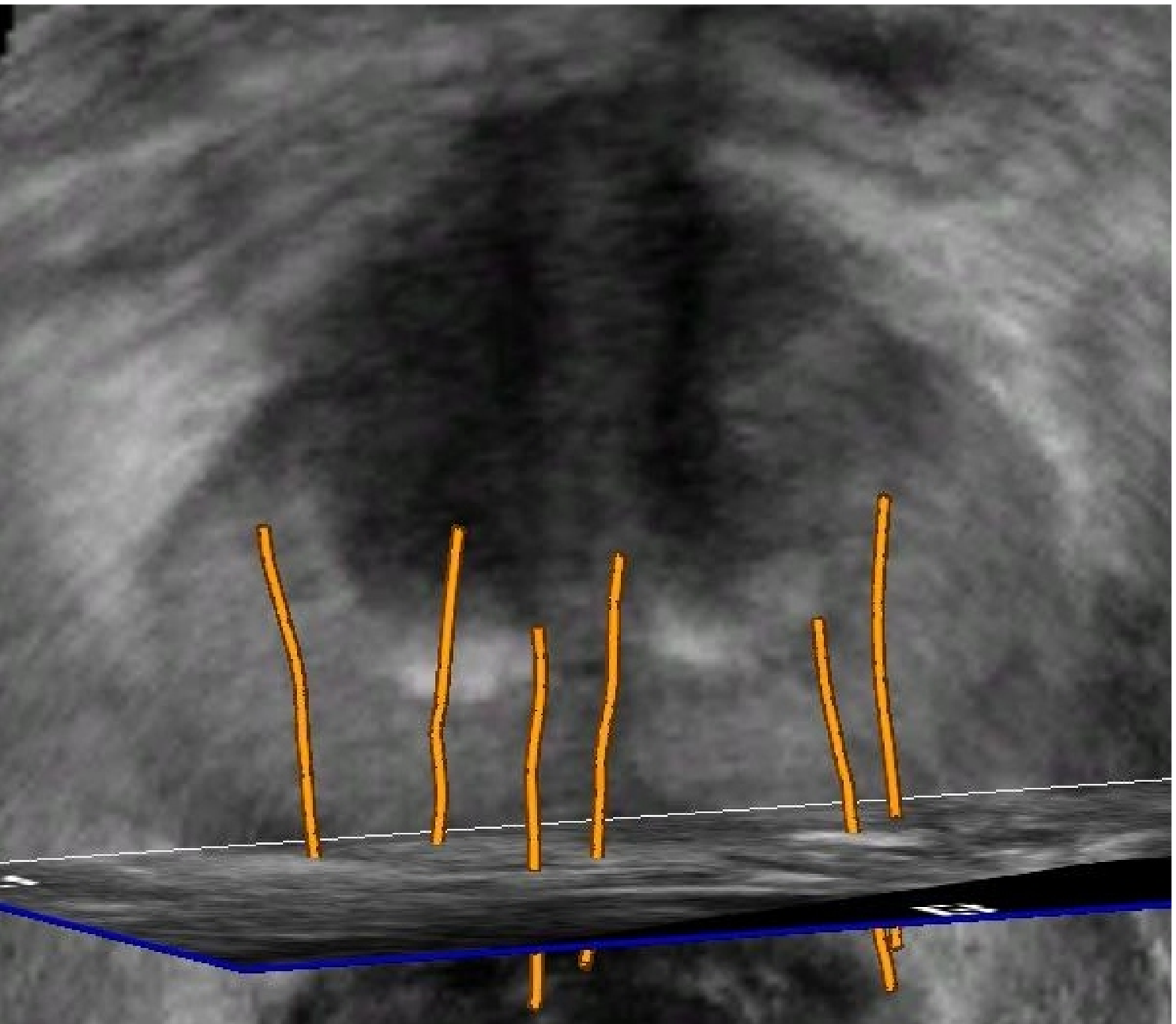}}\hspace{10pt}
\subfigure[]{\includegraphics[width=0.3204\textwidth]{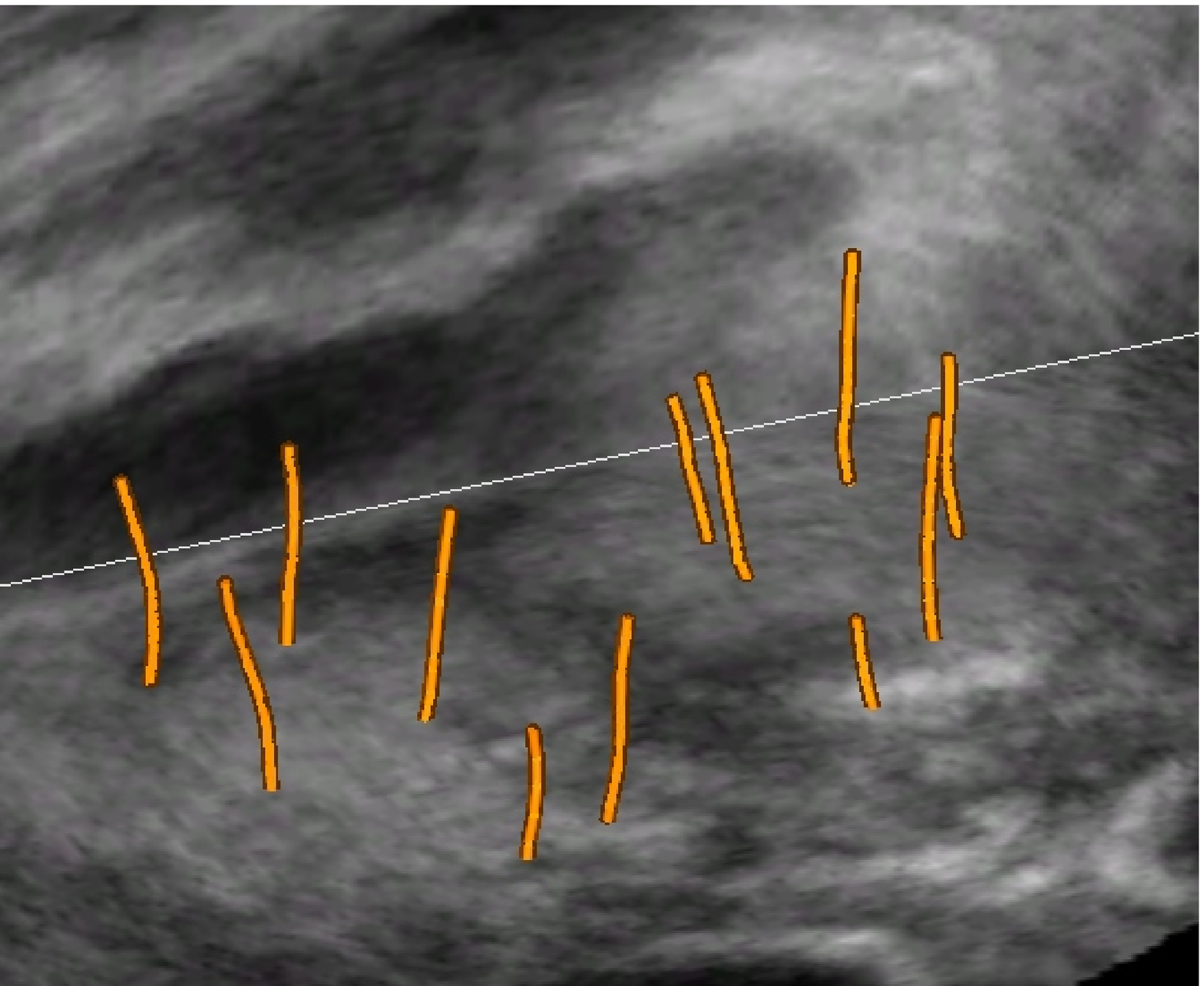}}
\caption{Tissue deformation corrected 3D biopsy maps in reference space.}
\label{fig:maps}
\end{figure}

The framework was validated on 278 registrations of 295 US volumes from 17 patients. The 17 reference images were acquired shortly before the intervention, and the tracking images were acquired after a biopsy shot. The clinical protocol was approved by the ethical committee of the XXX hospital, Town, Country, and all patients consented to participate to the study. The images were acquired with a GE Voluson and a RIC5-9 endorectal US probe. The algorithms were executed on a 4-core 2.6Ghz processor. In order to provide a reference gold standard for the evaluation of registration accuracy, experts manually segmented 467 point fiducials that were clearly identifiable on multiple images (e.g. calcifications and cysts). The distances between fiducial pairs were measured after registration to estimate the local accuracy. Note that the unavoidable segmentation error increases the measured error in average; this approach hence underestimates accuracy. Accuracy was computed for all registrations that were qualified as valid by experts after visual inspection, which represent 97,8$\%$ of the registrations. The results for both rigid and elastic registration are given in Tab.~\ref{nl:tab:accuracy}, and a visual illustration of the registration performance is given in Fig.~\ref{fig:SSD}. Fig.~\ref{fig:maps} shows 3D biopsy maps created with our biopsy tracking system.
\begin{table}
\centering
\begin{tabular}{@{\hspace{10pt}}r@{\hspace{10pt}}|@{\hspace{10pt}}c@{\hspace{10pt}}c@{\hspace{10pt}}c@{\hspace{10pt}}c@{\hspace{10pt}}}
	          & mean      & standard  & max      & execution \\
		      & distance  & deviation & distance & time (mean) \\[3pt] \hline \\[-7pt] 
unregistered  & 13.76 mm  & 7.89 mm   & 51.61 mm & - \\
rigid         & ~1.33 mm  & 0.85 mm   & ~4.19 mm & 2.1 s \\
elastic       & ~0.83 mm  & 0.54 mm   & ~4.14 mm & 6.8 s 
\end{tabular}
\caption{Accuracy study.}
\label{nl:tab:accuracy}
\end{table}

\section{Discussion and conclusion}
Deformation estimation achieves an overall accuracy of at least 0.83$\pm$0.54 mm on real patient data. This corresponds to an error reduction of 40$\%$ when compared to rigid 3D-3D registration.  The average computation time of the registration was only 6.8s. We are confident that the algorithm can be accelerated to below 1s on the same machine with simple optimization and parallelization techniques, which is sufficient for assisted needle placement. With specialized standard hardware (GPUs), at least 5Hz should be feasible. 

Biopsy tracking systems potentially add significant clinical value to prostate cancer diagnosis and therapy planning. Immediate advantages are the possibility to avoid resampling of already biopsied tissues when repeating a biopsy series, interventional quality control of the biopsy distribution (e.g. detection of unsampled areas) and computer-assisted guidance to non-systematic targets. The latter could for example be identified on MR/spectroMR images of the gland. Moreover, the improved knowledge about the biopsy and thus the cancer position could be used to implement focal therapy strategies for prostate cancer. 3D US based elastic tracking can provide the precision required for such therapeutic applications.

\bibliographystyle{splncs}
\bibliography{miccai09}

\end{document}